\documentclass[twocolumn,aps,prl,floatfix]{revtex4}
\usepackage{graphicx,amsmath,subfigure,epsfig,psfrag} %graphics
\usepackage{amsmath}
\usepackage{amssymb}

\providecommand{\xx}{\mathbf{x}}

\providecommand{\calH}{\mathcal{H}}
\providecommand{\mean}[1]{\left\langle #1 \right\rangle}

\begin{document}
\title{Using Inhomogeneity to Raise Superconducting Critical Temperatures}
\author{Y. L. Loh}
\author{E. W. Carlson}
\affiliation{Department of Physics, Purdue University, West Lafayette, IN  47907 }
\date{\today}
%%YL
%\date{Commenced 2006-3-28, \TeX ed \today{}}

\begin{abstract}
Superconductors with low superfluid density can be described by XY models.  
In such models the scale of the transition temperature $T_c$ is
largely set by the zero temperature phase stiffness (helicity modulus), a 
long-wavelength property of the system: $T_c = A \Upsilon(0)$.
However, the constant $A$ is a non-universal number, depending on dimensionality
and the degree of inhomogeneity.  In this Letter, we discuss strategies for maximizing 
$A$ for 2D XY models, that is, how to maximize the transition temperature with respect
to the zero temperature, long wavelength properties.  
We find that a framework type of inhomogeneity can increase the transition temperature significantly.
For comparison, we present similar results for Ising models.  
%We show that certain types of inhomogeneity can raise the $T_c$ of a 2D XY model representing a superconductor.
%We study the behavior of the helicity modulus in such cases and obtain an understanding how the shape of the $\Upsilon(T)$ curve changes....
\end{abstract}
\maketitle

%====================================================================
%INTRODUCTION
%====================================================================

Many strongly correlated models exhibit either local inhomogeneity (whether ordered or disordered)
or out-right phase separation,
%%EC
%\cite{frustrated,larged,largeJ,assaS,AssaN,spherical}
and there is experimental evidence that some degree of local electronic
inhomogeneity takes place in various parts of the phase diagrams of
transition metal oxides such as nickelates, cuprates, and 
manganites.\footnote{For recent reviews, see \onlinecite{dagotto-science,concepts}.}
It is important to understand how the macroscopic properties emerge
out of the mesoscale structure, and whether it
has detectable consequences for the observed phases, such
as the technologically important superconductivity observed in some
strongly correlated systems.  That is, does local inhomogeneity help or harm superconductivity, or is it a side issue entirely?

For superconductors with low superfluid  density,
the transition temperature is dominated by phase fluctuations of the
superconducting order parameter, 
and the transition may be captured by an XY model.  
Although the transition temperature in XY models is largely set by zero temperature,
long wavelength properties of the system, dimensionality and inhomogeneity
also play a role.  That is, $T_c = A \Upsilon(T=0)$ where $\Upsilon$ is the phase
stiffness or helicity modulus (proportional  to the superfluid density in a superconductor),
and $A$ is a non-universal number of order 1.
We focus here on how inhomogeneity may be used to maximize $A$,
and therefore enhance the transition temperature 
with respect to the zero temperature, long wavelength properties of the system,
as compared to the uniform case.  

It has generally been expected that inhomogeneity should decrease $T_c$, especially
to the extent that it introduces disorder or competing orders.  
However, this intuition has been violated  even in conventional superconductors
when mesoscale structures were introduced.  For example, many authors have reported that
%%EC temperature -> temperatures
the transition temperatures of 
%Al 
%(and other metals\footnote{\normalsize{This has been shown for other metals too, such as 
%In,\cite{grains-al-in,films-al-in-sn,films-many} Sn,\cite{films-al-in-sn,grains-sn,films-many}, and other soft metals.\cite{films-many} 
%Much of this effect has been attributed to phonon softening at 
%boundaries.\cite{inhomoSC-dickey1968,layers-al-cu-sn,inhomoSC-mcmillan}}}) 
Al, In, Sn, and other soft metals
can be increased over that of the bulk in the case of grains, films, or 
layered structures.\footnote{For a review, see \onlinecite{bose}.}
%grains,\cite{grains-al-in,grains-al} films,\cite{films-al-in-sn,films-many} or layered structures.\cite{layers-al-cu-sn,layers-zavaritsky,layers-kammerer}

Similar issues have been addressed theoretically in Hubbard models. 
For attractive models, spatial variation in $U$ can increase $T_c$\cite{PhysRevB.72.060502,PhysRevB.73.104518}
for checkerboard and stripe patterns, especially when the modulation wavelength is close to the
coherence length.\cite{PhysRevB.72.060502}  
It is not surprising that attractive-$U$ Hubbard models benefit from inhomogeneity.  
According to BCS theory, the pairing energy scale has a strong non-linear dependence on the attraction, $\Delta \sim T_c \sim e^{-1/\nu U}$.  Since $\mean{e^{-1/\nu U}} \geq e^{-1/\nu \mean{U}}$,  the local pairing amplifies favorable
spatial variations in $U$.  
Even in the repulsive case, it has been shown that the superconducting gap of coupled 2-leg ladder systems
is maximized for intermediate coupling between ladders.\cite{PhysRevB.68.180503}

%Arrigoni and Kivelson \cite{PhysRevB.68.180503} studied, analytically, a repulsive-$U$ Hubbard model on coupled two-leg ladders.  They found that the superconducting gap (according to their definition) is greatest for a certain optimal coupling between the ladders, that is, when the system is inhomogeneous.  \textbf{This is an interesting non-trivial result.}
%Martin and Kivelson \cite{PhysRevB.72.060502} studied an attractive-$U$ Hubbard model with an inhomogeneous, sinusoidally varying $U$.  They found that the superconducting $T_c$ was greatest when the wavelength of the variation was of the same order as the coherence length.
%Aryanpour \etal \cite{PhysRevB.73.104518} studied an attractive-$U$ 2D Hubbard model, where the attraction at each site $U_i$ was chosen randomly or according to checkerboard or stripe patterns, such the total attractive potential $\sum_i U_i$ was conserved.  They used ``Bogoliubov-de Gennes mean-field theory'' as well as a ``Monte Carlo mean field'' method.  They found that such inhomogeneity in $U_i$ often enhances $\Delta$ as well as $T_c$.

%%%%%%%%%%% FIGURE Inhomo patterns %%%%%%%%%%%%
\begin{figure}[tb]
\psfrag{Js}{$J_s$}
\psfrag{Jw}{$J_w$}
{\centering
\subfigure[1D modulation]
{\includegraphics[width=0.46\columnwidth]{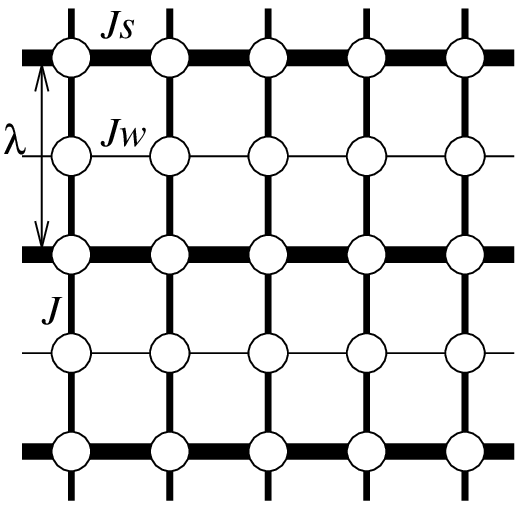}\label{fig:1dmodulation}}
\hspace{0.01\columnwidth}
\subfigure[2D modulation]
{\includegraphics[width=0.46\columnwidth]{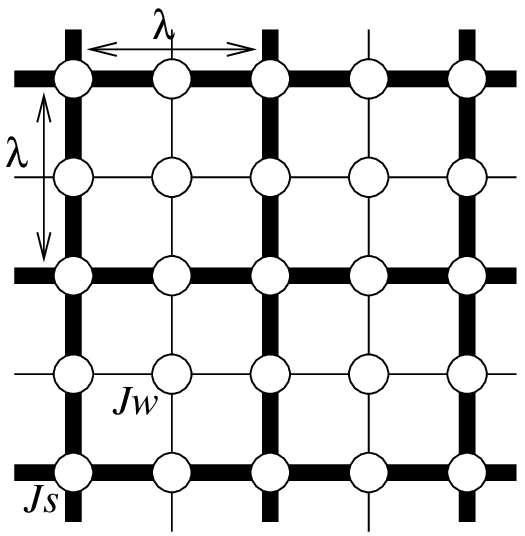}\label{fig:2dmodulation}}
\par}
\caption{In this Letter we study the above inhomogeneity patterns, where certain bonds are made stronger ($J_{s}$) and others weaker ($J_{w}$) such that the total coupling is preserved.
\label{fig:decoratedlattices}}
\end{figure}
%%%%%%%%%%%%%%%%%%%%%%%%%%%%%%%%%%%%

%====================================================================
%\subsection{Motivation and summary of this paper}
%====================================================================

In this Letter, we predict ways to maximize the bulk transition temperature with respect to the
zero temperature, long wavelength properties of the system.  
As a model for superconductors with low superfluid density, 
we study numerically 2D XY models with inhomogeneous couplings.
We are interested in patterns of the coupling constants that increase the transition temperature
over the homogeneous case.  In order to make fair comparisons, we require that the 
zero temperature, long wavelength properties of the system remain unchanged.
That is, we will not increase the low temperature energy density of the system,
or the low temperature helicity modulus.  
For comparison, we also study Ising models with inhomogeneous couplings; the results support our findings for XY models.
We find that although most patterns of inhomogeneity reduce the transition temperature $T_c$, there are indeed certain 
``framework'' patterns of inhomogeneity that increase $T_c$ by up to a theoretical maximum of 76\%.

%\textbf{Certain superconductors can be described by XY models.  In particular, the underdoped cuprates have very low superfluid density, and there is some reason to believe that their superconducting transition is phase-dominated, and can be well-captured by an XY-type model.}
%XY models can also be used to describe superconductors that are purposely manufactured to be granular, as well as Josephson junction arrays.

%%%%%%%%%%% FIGURE FSS  details %%%%%%%%%%%%%%
\begin{figure}[htb]
{\centering
\subfigure[Dimensionless helicity modulus $u=\Upsilon/T$ as a function of system size.  The curves are solutions of the scaling equations, Eq.~\eqref{e:ktrg}, chosen to fit the data (crosses).  The dashed line is $\Upsilon/T=\tfrac{2}{\pi}$.]
{\includegraphics[width=0.9\columnwidth]{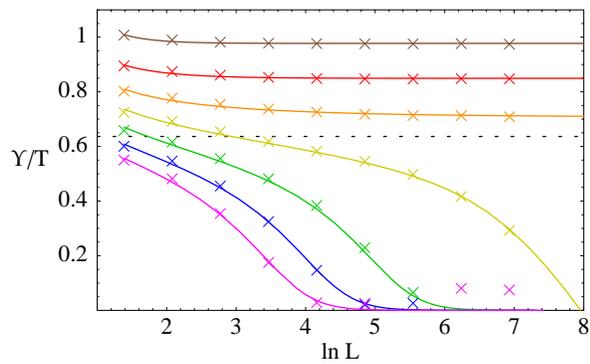}\label{fig:stiffnessfssdetails}}
\subfigure[Helicity modulus as a function of temperature.  The black curve is $\Upsilon(L=\infty,T)$ obtained by FSS.  The dashed line is $\Upsilon=\tfrac{2}{\pi}T$.]
{\includegraphics[width=0.9\columnwidth]{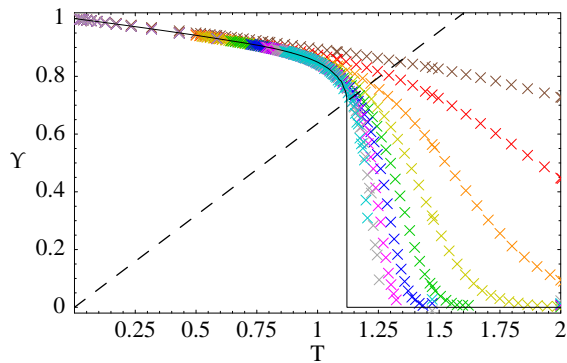}\label{fig:stiffnessfss}}
\par}
\caption{Finite-size scaling.  The crosses are Monte Carlo results for the helicity modulus $\Upsilon(L,T)$ of a 4x4-modulated XY model with $J_w=0$ and 
%%EC times J_avg so it has units of energy
$J_s=4J_{\rm avg}$, for $T=0.9, 1.0, ..., 1.5$ in units of $J_{\rm avg}$
and $L=4,8,...1024$.
}
\end{figure}
%%%%%%%%%%%%%%%%%%%%%%%%%%%%%%%%%%%%

The XY model has the following classical Hamiltonian:
	\begin{align}
	\calH_\text{XY} [\theta] &=
		-\sum_{\mean{ij}} J_{ij} \cos (\theta_i - \theta_j)
	\end{align}
where $i,j$ are site labels, $J_{ij}$ are nearest-neighbour couplings, and $\theta_\xx$ are real-valued phase (angle) variables.
We choose the  couplings $J_{ij}$ so as to preserve $\sum_{ij} J_{ij}$.  This leaves the zero-temperature properties of the system unchanged, such as the helicity modulus $\Upsilon(T=0)$ and the energy $U(T=0)$.   
We consider here only two-dimensional models.

We are interested in how to optimize the transition temperature, and to study this 
we focus on the behavior of the helicity modulus $\Upsilon(T)$ of the XY model, which is directly proportional to the superfluid stiffness $n_s$ of a phase-dominated superconductor.
%%YL: added ref to schultka-manousakis
\footnote{The helicity modulus measures the change in the free energy caused by a small change in the phase angle \cite{schultka-manousakis}.
} by 
%, which can be measured experimentally through the  London penetration depth.
We study this quantity via Monte Carlo simulations on square lattices, where 
$\Upsilon(T)$ can be calculated using
	\begin{align}
	\Upsilon
	&=
	\frac{1}{2V} 
	\mean{ 
		\sum_{\mean{ij}} J_{ij} \cos (\theta_i - \theta_j)
	-	\beta 	\bigg[
		\sum_{\mean{ij}} J_{ij} \sin (\theta_i - \theta_j)
		\bigg]^2
	}
	.
	\end{align}
We use the Wolff cluster algorithm\cite{wolff1989}, which is the fastest serial algorithm for our purposes.  The $\theta$ variables are stored and manipulated as two-vectors to avoid trigonometric function calls.  
%%EC Can we do without this footnote?
%\footnote{
%Simulation of the dual model, the solid-on-solid (SOS) model, would allow us to exploit the series reduction formulas and would be somewhat more efficient, but this approach would only work for the 2D XY model (and not for the other models which we studied, which we will present in another paper).
%}

%EC -- Not sure we need all the detail of the next few sentences.  
%The 2D XY model is at its lower critical dimension and fluctuations are severe.
%The transition in the 2D XY model is of the Kosterlitz-Thouless type; the correlation length grows exponentially as one approaches the KT transition; therefore, simulations suffer from large finite-size errors.
%To get a sufficiently accurate and unbiased estimate of $T_c$, it is necessary to perform finite-size scaling (FSS) to eliminate/compensate for finite-size errors. [{\em The previous sentence goes without saying -- probably don't need to say it. -EC}]
%The easiest type of FSS is using the intersection point of the Binder parameter curves, but this method may not be applicable to KT transitions.
%The logarithmic nature of the scaling makes it {\bf difficult or impossible} [{\em Do we want to say that? -EC}] to collapse the data onto a universal scaling curve. 

In order to obtain reliable estimates of $T_c$, we have performed finite-size scaling (FSS) on $\Upsilon(L,T)$ in the following manner.
%%YL: added ref to schultka-manousakis
The KT transition can be described by a two-parameter scaling flow \cite{kosterlitz1974,nelson1977,jose1977,schultka-manousakis} for the dimensionless helicity modulus $u=\Upsilon/T$ and the `vortex fugacity' $y$, 
	\begin{align}
	\frac{du}{dl} &= -4\pi^3 u^2 y^2,
	\quad
	\frac{dy}{dl} = (2-\pi u)y
	,
	\label{e:ktrg}
	\end{align}
where $l=\ln L$ is the length scale.  This pair of differential equations can be solved numerically, given initial values $u(l_0)=u_0$ and $y(l_0)=y_0$ (where $l_0$ is some reference length scale).  
%%EC 
%For each temperature $T$, we find $u_0$ and $y_0$ such that the function $u(l)$ fits the Monte Carlo data $u(\ln L)$ very well for the available system sizes, 
For each temperature $T$, we choose $u_0$ and $y_0$ 
so as to obtain a good fit of $u(l)$ to the Monte Carlo data for the available system sizes, 
$4 \leq L \leq 1024$ (see Fig.~\ref{fig:stiffnessfssdetails}).  
We then integrate the differential equations all the way to $l=\infty$.  This gives $u(\infty)$ and hence the helicity modulus in the infinite-size limit $\Upsilon(T,L=\infty)$, shown in Fig.~\ref{fig:stiffnessfss}.
\footnote{This treatment neglects the tiny corrections due to non-zero winding numbers\cite{hasenbusch2005}, but it is adequate for the level of accuracy of the present work.
As a check, we have estimated $T_c$ using other methods, such as finite-size scaling for the susceptibility $\chi(L,T)$ and for the Wolff cluster size $N_\text{clust}(L,T)$, based on the predictions of Kosterlitz-Thouless theory that $\chi(L,T_c) \sim L^{7/4} (\ln L)^{1/8}$ and that $\chi(\infty,T-T_c) \sim t^{-1/16} e^{ct^{-1/2}}$.  The results agree with those obtained from finite-size scaling for $\Upsilon$.
}

\begin{figure}[htb]
\includegraphics[width=0.9\columnwidth]{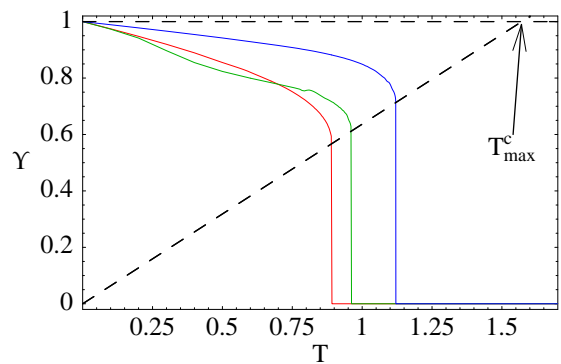}
\caption{$\Upsilon(T,L=\infty)$ for 2D XY models.  The red curve is for the homogeneous 2D XY model.  The green curve is for a 4x4 modulation with $J_\text{strong}=3.4$ and $J_\text{weak}=0.2$.  The blue curve is for a 4x4 modulation with $J_\text{strong}=4$ and $J_\text{weak}=0.$
 The dashed line is $\Upsilon=\frac{2}{\pi}T$; the arrow indicates a theoretical upper bound on $T_c$, 
 %%EC times J_avg, so it has units of energy. 
 $T^c_\text{max}=\frac{\pi}{2}J_{\rm avg}$.
}
\label{fig:knees}
\end{figure}
%%%%%%%%%%%%%%%%%%%%%%%%%%%%%%%%%%%%

Our most important result is that \emph{by redistributing the bond strengths of an XY model in certain inhomogeneous patterns, it is possible to increase $T_c$.} 
As a concrete example of how this comes about,
we show how the shape of the helicity modulus curve
{\em vs.} temperature is changed by introducing inhomogeneity. 
In Fig.~\ref{fig:knees}, we show our simulations of $\Upsilon(T)$,
extrapolated to infinite system size $L \rightarrow \infty$,
for a 2D inhomogeneity of the type shown in Fig.~\ref{fig:2dmodulation},
using $\lambda = 4 a$ where $a$ is the underlying  lattice constant.  
In the blue and green curves of Fig.~\ref{fig:knees}, the coupling constant $J_{ij}$ has been made stronger 
on the darker lines in Fig.~\ref{fig:2dmodulation},  ($J_{\rm strong}$), and weaker on the lighter lines ($J_{\rm weak}$).
We compare these to the uniform case (the red curve) with $J$
set equal to the spatial average of $J_{\rm weak}$ and $J_{\rm strong}$,
$J = J_{\rm avg} \equiv (J_{\rm strong} + (\lambda -1)J_{\rm weak})/\lambda$.
Thus for all curves shown in Fig.~\ref{fig:knees}, 
the zero temperature helicity modulus $\Upsilon(T=0)$ and the 
zero temperature free energy are the same.  

%%YL: added ref to schultka-manousakis
In the uniform case, it is known that $T_c = 0.8929J$ \cite{schultka-manousakis,hasenbusch2005},
 and that the low temperature slope of the helicity modulus $\Upsilon^{'}(0) = 1/4$ \cite{classphs,roddick-stroud}.
The green curve in Fig.~\ref{fig:knees} shows the helicity modulus for  $J_{\rm strong} =  3.4$
and $J_{\rm weak} = 0.2$.  In this case, 
the transition temperature is enhanced by $8\%$ above
the homogeneous case.  For the case of extreme inhomogeneity with
$J_{\rm strong} = 4$ and $J_{\rm weak} = 0$ (the blue curve),
the transition temperature is $25\%$ higher than in the uniform case.
The shape of the green curve demonstrates the separation of energy
scales that happens with inhomogeneity.  Notice that at the very lowest temperatures,
the green curve is dominated by the long-wavelength average of the coupling constants,
and the low temperature linear slope of the helicity modulus is identical to that of the 
homogeneous case.  As temperature is raised, the slope increases in magnitude,
as the weak plaquettes become disordered.  Then, at a higher temperature, the slope 
approaches that of the blue curve, indicating that at higher temperatures the 
helicity modulus is dominated by $J_{\rm strong}$.  It is this shallower high temperature
slope which causes the helicity modulus to overshoot the homogenous $T_c$,
and leads to an inhomogeneity-induced enhancement of the transition temperature.

\begin{figure}[htb]
{\centering
\subfigure
{\resizebox*{0.98\columnwidth}{!}{\includegraphics{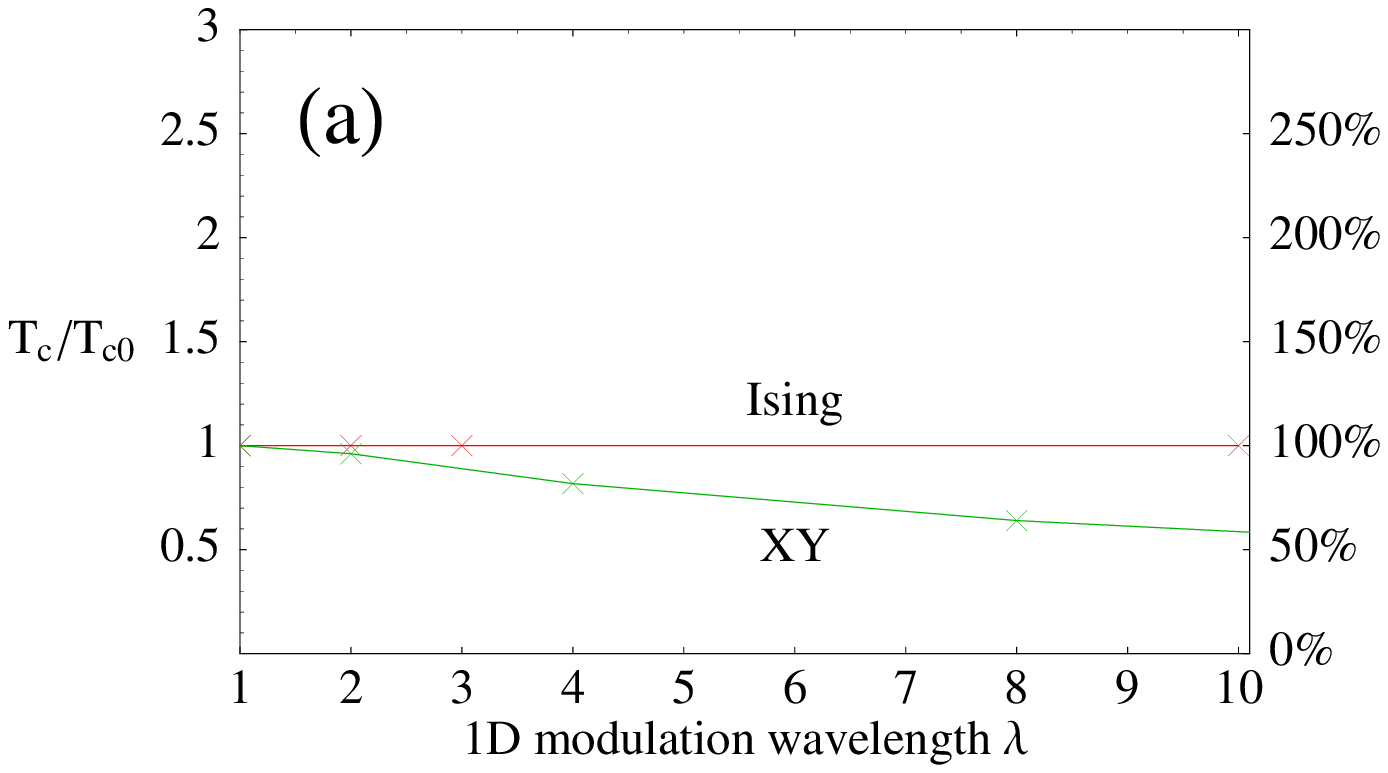}\label{fig:1dmodtc}}} 
\subfigure
{\resizebox*{0.98\columnwidth}{!}{\includegraphics{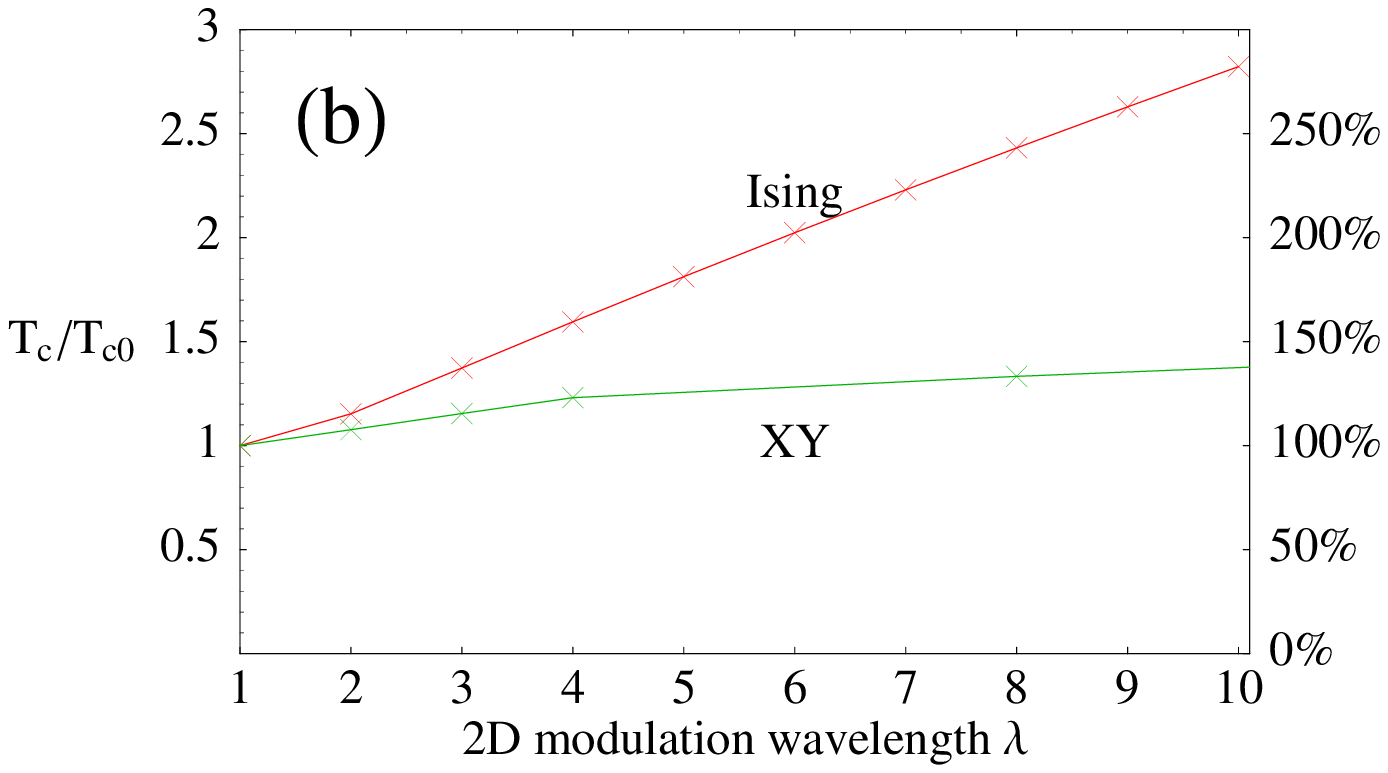}\label{fig:2dmodtc}}} 
\par}
\caption{Critical temperatures $T_c$ of the lattices depicted in Fig.~\ref{fig:1dmodulation} and \ref{fig:2dmodulation}, as a function of the wavelength of the inhomogeneity.
\label{fig:decoratedtc}}
\end{figure}

For 2D patterns like those in Fig.~\ref{fig:2dmodulation},
the enhancement of the transition temperature increases with $\lambda$,
as shown by the green curve of Fig.~\ref{fig:2dmodtc}.
However, the enhancement is constrained by the
zero temperature helicity modulus.  
Even in the presence of inhomogeneity, in 2D the system remains in the KT universality class, 
and the helicity modulus has a universal jump at the transition such that $\Upsilon(T_c)=0.6365 T_c$.  
Since thermal fluctuations introduce disorder, $\Upsilon(T_c) < \Upsilon(T=0)$,
so that $T_c/\Upsilon(0) \leq 1.57$, 
or equivalently, $T_c/T_{c0} \leq 1.76$.  
This theoretical upper bound on $T_c$ is illustrated in Fig.~\ref{fig:knees}.
%Fig.~\ref{fig:2dmodtc} confirms that $T_c$ never increases past this theoretical upper bound
%as the modulation wavelength $\lambda$ increases.  
That is, although the zero temperature properties of the system may be used 
as a predictor of the transition temperature, $T_c = A \Upsilon(T=0)$,
the constant $A$ is non-universal.
In fact, $A=T_c/\Upsilon(0)$ may be useful for \emph{characterizing} the degree of inhomogeneity:
it increases from $0.89$ for the uniform XY model up to a theoretical maximum of $1.57$.  For a 2D system, a large measured value of this ratio may indicate substantial inhomogeneity.  (An increase in this ratio may also indicate higher dimensionality.\cite{classphs})

We have also considered one-dimensional modulations, like those in Fig.~\ref{fig:1dmodulation}.
Since this type of inhomogeneity drives the system towards more one-dimensional physics
where a phase transition is forbidden by the Mermin-Wagner theorem,
the transition temperature decreases, as shown in Fig.~\ref{fig:1dmodtc}. 
Hence the enhancement of $T_c$ is not additive --- the effect of a 2D modulation is \emph{not} double that of a 1D modulation.

%In order to increase the transition temperature, the strong bonds must be aligned with each other in a framework structure that extends in all directions (as in Fig.~\ref{fig:2dmodulation}).
%That is, the bond strengths must have a 2-dimensional modulation.  A 1-dimensional modulation
%(Fig.~\ref{fig:1dmodulation}) does not affect the Ising $T_c$, and it moreover decreases the $T_c$ of the XY model.  See Figs.~\ref{fig:1dmodtc} and \ref{fig:2dmodtc}.
%This is interesting because it suggests that the enhancement of $T_c$ is not additive or perturbative: the effect of a $2d$ modulation is \emph{not} double the effect of a $1d$ modulation!
%[{\em This is probably mostly a dimensional effect.  1D is below the lower critical dimension, and there is no transition. --EC}]

Fig.~\ref{fig:decoratedtc} also shows the effect of inhomogeneity in the Ising model,
for comparison.  Inhomogeneous Ising  models may be described by the Hamiltonian
	\begin{align}
	\calH_\text{Ising}[\sigma]	
	&=-\sum_{\mean{ij}} J_{ij} \sigma_i \sigma_j, \quad \sigma_i=\pm 1.
	\end{align}
%The critical temperature $T_c$ of a 2D Ising model with periodic inhomogeneity, such as in Fig.~\ref{fig:decoratedlattices}, can be determined using the Pfaffian method\cite{fisher1966} with no need for finite-size scaling.
As with the XY model, we restrict ourselves to two dimensions.
For the purpose of studying different length scales of inhomogeneity, 
we focus on an extreme type of inhomogeneity with $J_{\rm weak}=0$, and
$J_{\rm strong} = \lambda J_{\rm avg}$.  
%, which can be treated even more easily.  
Such patterns correspond to `decorated' lattices.  By integrating out all doubly-coordinated spins (that is, by applying the so-called decorated-iteration transformation, $J_\text{eff} = \tanh^{-1} (\tanh J_1 \tanh J_2)$), one can reduce a decorated lattice to a primitive lattice and thus obtain an exact expression for its $T_c$ (Eq.~\eqref{eqn:2ddecoratedisingtc}).
%%EC 
%\footnote{For more complicated patterns not amenable to decoration-iteration, the Pfaffian method\cite{fisher1966} can be used to calculate $T_c$.  Had we been considering random inhomogeneity, the recently developed bond-propagation algorithm\cite{loh2006} would be our method of choice.}
\footnote{For more complicated patterns not amenable to decoration-iteration,
$T_c$ can be calculated using either the Pfaffian method\cite{fisher1966}
({\em e.g.}, for periodic inhomogeneity), or the recently developed bond-propagation algorithm\cite{loh2006} (which is the most efficient in the case of bond dilution).}
(Unfortunately, the decoration-iteration transformation for XY models involves an infinite set of Fourier components of the potential\cite{jose1977} and it does not lead to exact results for $T_c$.) 
%YL: Jose's paper uses  Migdal's bond-moving approximate RG scheme to get T_KT~0.8

%, that is, by using the series reduction formulas for Ising  models.
%\footnote{These lattices resemble those generated in the approximate Migdal-Kadanoff bond-moving renormalization scheme, but here we take those lattices as the starting point, so there is no approximation involved.}  %YL: this is not necessary, because our paper isn't focusing on Ising models, and if people are stupid enough to confuse us with migdal-kadanoff, it's not our problme
%For two bonds in parallel, $J_{12}=J_1+J_2$.
%In the case of the XY model, each bond is described by a coupling \emph{function} $J(\theta)$, which is equal to $J \cos \theta$ for the standard XY model, where $\theta$ is the `phase difference' between adjacent sites.  In the dual representation we have an infinite array of Fourier series coefficients, $K_m=\ln \int_0^{2\pi} d\theta~ e^{im\theta} e^{J(\theta)}$.  To combine bonds in parallel, we write $J_{12} (\theta) \equiv J_1(\theta) + J_2(\theta)$.  To combine bonds in series, we write $K^{12}_m \equiv K^1_m + K^2_m$.
%These formulas are nonlinear, unlike the corresponding formulas for resistor networks, and lead to more interesting behavior.
%\footnote{Warning: you can only define a planar dual in 2D, and some of our systems are 3D; but the equations stil hold.}

In Fig.~\ref{fig:decoratedtc}, we show the effect of extreme inhomogeneity 
({\em i.e.}, with $J_{\rm weak}=0$) on the transition temperatures in Ising
and XY models.  We use the maximum value of $J_{\rm strong}=\lambda J_{\rm avg}$, because
for a given  wavelength $\lambda$, this gives the largest enhancement of $T_c$
while conserving the average coupling $J_{\rm avg}$.
While we are interested primarily in superconductors with small superfluid density,
which can be captured with an XY model, we also show results for
the Ising model, for which results can be obtained analytically as described above.
Fig.~\ref{fig:1dmodtc} shows the effect of a purely 1D modulation,
as a function of distance $\lambda$ between strong bonds $J_{\rm strong}$,
chosen so as to preserve the zero temperature, long wavelength properties 
of the system.  The pattern of coupling constants is shown in Fig.~\ref{fig:1dmodulation}.
In the Ising case, the transition temperature is unchanged
by this procedure.  In the XY case, the transition temperature decreases monotonically
with $\lambda$.

The effect of a 2D modulation is shown in Fig.~\ref{fig:2dmodtc}.
Again, parameters are chosen so as to preserve the zero temperature 
properties of the system.  Fig.~\ref{fig:2dmodulation} shows the pattern
of coupling constants.  Here, the transition temperature in the Ising case increases
as 
\begin{equation}
T_c = \frac{2  \lambda J_{\rm avg} }{ \sinh^{-1} [{\rm csch} ( \frac{1}{\lambda} \sinh^{-1} 1) ]}~.
\label{eqn:2ddecoratedisingtc}
\end{equation}
%The following is the expression for Tc of a decorated Ising model, with "wavelength" lambda in both directions, and with Jstrong=lambda*J0, Jweak=0:
%where ${\rm arcsinh}(1) = 0.881374$.  
One of the occurrences of lambda in this equation is due to taking   $\lambda$ bonds in parallel, to form``bundles'', and the other occurrence is due to taking  $\lambda$  bundles in series.
For XY models, the transition temperature also increases monotonically with
modulation length $\lambda$.  In this case there is an upper bound, set by
the zero temperature properties of the system, as shown in Fig.~\ref{fig:knees}.
That is, the maximum enhancement of $T_c$ possible with this type of inhomogeneity
in an XY model is $76\%$.  

%====================================================================
%DISCUSSION AND CONCLUSIONS
%====================================================================

In conclusion, we have shown that certain types of inhomogeneity 
can increase the transition temperature of Ising and XY models.
Specifically, two-dimensional modulations of the coupling constants
that preserve the spatial average coupling increase the transition temperature
over that of the uniform case. 
One-dimensional modulations depress the transition in XY models,
and leave the transition temperature unchanged in Ising models.  
Our results for 2D XY models may indicate that certain types of inhomogeneity
can result in an enhancement of 
superconductivity in systems with low superfluid density.

%%%%%%%%%%%%%%%%%%%%%%%%%%%%%%%%%%%%

%====================================================================
%ACKNOWLEDGMENTS
%====================================================================

It is a pleasure to thank S. A. Kivelson, E. Manousakis, and D. Stroud for 
%%EC added helpful
helpful discussions.
This work was supported by Purdue University (YLL), Research Corporation (YLL), 
and by the Purdue Research Foundation (EWC).  EWC is a Cottrell Scholar of Research Corporation.
This research was supported in part
%%YL
%by (agreement #*************) 
through 
computing resources provided by Information Technology at Purdue-the Rosen Center for Advanced Computing, West Lafayette, Indiana.

\bibliographystyle{forprl}
\bibliography{sc,rbim}

\begin{thebibliography}{10}

\bibitem{PhysRevB.72.060502}
I.~Martin, D.~Podolsky, and S.~A. Kivelson, {\it Phys. Rev. B\/}, {\bf 72},
  060502(R) (2005).

\bibitem{PhysRevB.73.104518}
K.~Aryanpour, {\it et~al.\/}, {\it Phys. Rev. B\/}, {\bf 73}, 104518 (2006).

\bibitem{PhysRevB.68.180503}
E.~Arrigoni and S.~A. Kivelson, {\it Phys. Rev. B\/}, {\bf 68}, 180503(R)
  (2003).

\bibitem{wolff1989}
U.~Wolff, {\it Phys. Rev. Lett.\/}, {\bf 62}, 361 (1989).

\bibitem{kosterlitz1974}
J.~M. Kosterlitz, {\it Journal of Physics C: Solid State Physics\/}, {\bf 7},
  1046 (1974).

\bibitem{nelson1977}
D.~R. Nelson and J.~M. Kosterlitz, {\it Phys. Rev. Lett.\/}, {\bf 39}, 1201
  (1977).

\bibitem{jose1977}
J.~V. Jos\'e, L.~P. Kadanoff, S.~Kirkpatrick, and D.~R. Nelson, {\it Phys. Rev.
  B\/}, {\bf 16}, 1217 (1977).

\bibitem{schultka-manousakis}
N.~Schultka and E.~Manousakis, {\it Phys. Rev. B\/}, {\bf 49}, 12071 (1994).

\bibitem{hasenbusch2005}
U.~Wolff, {\it J. Phys. A\/}, {\bf 38}, 5869 (2005).

\bibitem{classphs}
E.~W. Carlson, S.~A. Kivelson, V.~J. Emery, and E.~Manousakis, {\it Phys. Rev.
  Lett.\/}, {\bf 83}, 612 (1999).

\bibitem{roddick-stroud}
E.~Roddick and D.~Stroud, {\it Phys. Rev. Lett.\/}, {\bf 74}, 1430 (1995),
  linear-T dependence of SF density.

\bibitem{dagotto-science}
E.~Dagotto, {\it Science\/}, {\bf 309}, 257 (2005).

\bibitem{concepts}
E.~W. Carlson, V.~J. Emery, S.~A. Kivelson, and D.~Orgad, {\it Concepts in High
  Temperature Superconductivity\/}, Springer-Verlag (2004), in The Physics of
  Superconductors, Vol. II, ed. J.~Ketterson and K.~Benneman.

\bibitem{bose}
{S. Bose, P. Raychaudhuri, R. Banerjee, P. Vasa, and P. Ayyub}, {\it Phys. Rev.
  Lett.\/}, {\bf 95}, 147003 (2005).

\bibitem{fisher1966}
M.~E. Fisher, {\bf 7}, 10 (1966).

\bibitem{loh2006}
Y.~L. Loh and E.~W. Carlson, accepted by Phys. Rev. Lett. (2006).

\end{thebibliography}
\end{document}